\documentclass[secnumarabic,aps,prd,floatfix,nofootinbib,showkeys,twocolumn,showpacs,10pt,superscriptaddress]
{revtex4-1}
\usepackage{epsfig}
\usepackage{amsmath}

\begin{document}
\begin{titlepage}

\begin{center}
{\Large \bf Tolman-Ehrenfest-Klein Law in non-Riemannian geometries} 
\vskip 0.1cm

J. A. S. Lima$^{a,}$\footnote{jas.lima@iag.usp.br}, J. Santos$^{b,}$\footnote{janilo.santos@gmail.com}
\end{center}

\begin{quote}
$^a$Departamento de Astronomia, Universidade de S\~ao Paulo (USP)\\ Rua do
Mat\~ao 1226, 05508-900, S\~ao Paulo, SP, Brazil\\
$^b$Departamento de F\'isica Te\'orica e Experimental, Universidade Federal do Rio Grande do Norte (UFRN), 59000-072, Natal, RN, Brazil
\end{quote}

\centerline{\bf ABSTRACT}
\bigskip

Heat always flows from hotter to a colder temperature until thermal equilibrium  be finally restored in agreement with the usual (zeroth, first and second) laws of thermodynamics.  However, Tolman and Ehrenfest demonstrated that the relation between inertia and weight uniting all forms of energy in the framework of general relativity implies that the standard equilibrium  condition is violated in order to maintain the validity of the first and second law of thermodynamics. Here we demonstrate that the thermal equilibrium condition for a static self-gravitating fluid, besides being violated, is also heavily dependent on the underlying  spacetime geometry (whether Riemannian or non-Riemannian).  As a particular example, a new  equilibrium condition is deduced for a large class of Weyl and  $f(R)$ type gravity theories. Such results suggest that experiments based on the foundations of the heat theory (thermal sector) may also be used for confronting gravity theories and prospect the intrinsic geometric nature of the spacetime structure.

\bigskip
\end{titlepage}
\pagestyle{plain} 

\vskip 0.3cm
\bigskip





In 1930, Tolman \cite{T30}, and Tolman and Ehrenfest \cite{TE30} argued that
heat, as any other source of energy, would be subjected to gravity, and, as a result, a new thermo-gravitational effect absent in the classical thermodynamics was derived for a self-gravitating fluid described  by the general relativity. By  assuming a static  fluid configuration described by the line element (in our units $c=1$)
\begin{equation} \label{M}
ds^2\equiv g_{\alpha\beta}dx^{\alpha}dx^{\beta} = g_{00}dt^2 - g_{ij}dx^{i}dx^{j}\,,
\end{equation}
where all the $g_{\alpha\beta}(x^{i})$ are independent of time but
depends on the spatial coordinates $x^{i}\,\,
(i=1,2,3)$, they obtained an extended thermal equilibrium condition, sometimes dubbed Tolman-Ehrenfest (TE) law:
\begin{equation}\label{TE}
\partial_i\ln{T}=-{\partial}_i\ln\sqrt{g_{00}}\,\, \Leftrightarrow \,\,T\sqrt{g_{00}}= {\tilde T} = const.,
\end{equation} 
where $\partial_i \equiv \partial/\partial x^i$ and  ${T}$ is the temperature measured at different positions by local observers at rest in the single-fluid  system. This temperature T is the physical fundamental quantity that we mean by temperature at a given point of the medium. 

The most striking feature of such a relation is that the proper temperature varies from point to point within the self-gravitating fluid, a condition clearly  violating the so-called ``zeroth" law of thermodynamics. This happens even assuming  the energy conservation (first law) and also that the second law is strictly obeyed since the heat flow is absent \cite{Tolman34,S93}.  


In 1949, by assuming the validity of the TE law, it was proved by Klein \cite{Klein49}  that a similar relation is also valid for the chemical potential $\mu$, namely: 
\begin{equation}\label{KR}
\partial_i\ln{\mu} =-\partial_i\ln\sqrt{g_{00}}\,\,\,   \Leftrightarrow \,\,\,\mu\sqrt{g_{00}} =constant\,.
\end{equation} 
In the literature  (\ref{TE}) and (\ref{KR}) are thought as being independent relations, and, as such, were critically discussed by many authors based on different considerations~\cite{B1949,R2011,R2013,BH2016,S2018}. 


Nevertheless,  based  only on thermodynamics and general relativity theory (GRT), it was recently claimed that the original Tolman-Ehrenfest and Klein laws are not independent \cite{LAP2019}.  For the perfect fluid (source of curvature), it was found that the temperature, its chemical potential, and the metric coefficient $g_{00}(x^{i})$  are combined in such a way that a unique general relation uniting such quantities is obeyed  
\begin{equation}\label{TEKGRT}
\partial_i \ln (T\sqrt{g_{00}}) + \frac{\mu}{T\sigma}\,\partial_i \ln (\mu\sqrt{g_{00}}) = 0\,,
\end{equation}
where $\sigma$ is the specific entropy (per particle). 

In the above  expression, all local quantities depend only on the spatial position.  This relation uniting $T$, $\mu$ and $g_{00}$ holds regardless of the equation of state satisfied by the medium and also is fully independent of the  Einstein field equations. Henceforth, the above expression will be referred to as Tolman-Ehrenfest-Klein (TEK) law.  It means that the original TE law (\ref{TE}), as deduced in GRT, is valid only if the Klein law  $\mu\sqrt{g_{00}}=constant$ is obeyed. In particular, it remains valid for $\mu=0$.  Conversely,  Klein's relation (\ref{KR}) is also recovered when the TE law is assumed. However, the laws are not independent in the general case.    

Although independent of the Einstein field equations, it is worth  noticing that TEK type laws may also depend on the underlying theory of gravity. This is an interesting connection since the geometrical settings of Einstein's theory of gravitation is completely based on Riemannian geometry and further developments of GRT gave rise to new geometrical (non-Riemannian) spacetime structures for describing gravitational phenomena~\cite{Weyl}. Hence, it seems natural to investigate whether the  thermo-gravitational equilibrium condition depends on the spacetime sctructure.

In this concern, it is worth notice that there has been a renewed interest in the Weyl geometry~\cite{Scholz-I}, mainly due to the possibility of incorporating local conformal symmetry - also known as Weyl symmetry - into fundamental physics aiming to recast the standard model of particle physics plus gravity into a locally gauge invariant theory. This is done by adding new fields and introducing compensating gauge symmetries such that all couplings are dimensionless~\cite{Steinhardt,Moffat,Coumbe}. A natural framework in which such scale-invariance is realised is given by Weyl geometry~\cite{Scholz-II}. Some authors~\cite{Romero-Joel} have also shown how to formulate GRT using a particular Weylian geometry known as Weyl integrable spacetime (WIST), in which the gravitational field is described by two geometrical objects: the metric tensor and the so-called Weyl scalar field. We will follow this formulation in order to show how TEK's law presents itself in this geometry (see also \cite{salim}) .  

Another interesting extension of general relativity is known as $f(R)$ modified gravity. In such theories,  the standard GRT hypothesis of a strictly linear gravitational action in the Ricci curvature scalar $R$ is relaxed \cite{fR2,fR3}.  One aspect of this extension, known as Einstein-Palatini's approach (see \cite{fR4} for a review), considers the metric and the connections as basic variables in the gravitational action $S_g = S[g_{\mu\nu},\Gamma^{\lambda}_{\alpha\beta},...]$ thereby leading to non-Riemannian geometries~\cite{Sotiriou-Liberati}.
In this context, it is interesting both from methodological and theoretical viewpoints to investigate the TEK thermal-gravitational effect in extended theories of gravity, especially those formulated in a non-Riemannian environment. As we shall see, the results discussed here suggest naturally a new intriguing and compelling possibility, namely: the spacetime structure would be prospected through local experiments in the thermal sector guided by an extended TEK law. 

 With that goal in mind, we recall that the basics of relativistic thermodynamics is described by three fluxes: the energy-momentum tensor (EMT) $T^{\alpha\beta}$, the particle flux $N^{\alpha}$, and entropy flux $S^{\alpha}$. For a perfect fluid such quantities are defined by \cite{Degroot80} 

\begin{equation}\label{E1}
T^{\alpha\,\beta} = (\rho + p)u^{\alpha}u^{\beta} - pg^{\alpha\beta},\,\,N^{\alpha} = nu^{\alpha},\,\, S^{\alpha} = n\sigma u^{\alpha},  
\end{equation}
where $\rho$, $p$, $n$ and $\sigma$ denote
the energy density, thermostatic pressure, particle density
and specific entropy (per particle). Such quantities are related by the
Gibbs law \cite{SW72,SLC2002}
\begin{equation} \label{GL}
nTd\sigma = d\rho - \frac{\rho + p}{n}\,dn\,. 
\end{equation}
For a static inhomogeneous configuration as the one geometrically described by the metric (\ref{M}), such differentials are increments relating to neighbouring points. This means that for a static medium the above expression implies that
\begin{equation} \label{GL1}
nT\partial_i\sigma  =  \partial_i\rho  - \left(\frac{\rho + p}{n}\right)\partial_i n\,, 
\end{equation}
while the chemical potential is defined by the standard local form of Euler's relation:
\begin{equation}\label{mu}
\mu = \frac{\rho + p}{n} - T\sigma\,.
\end{equation}
The phenomenological expressions (\ref{E1})-(\ref{mu}) are basic consequences of the Einstein equivalence principle (EEP), and, as such, play a fundamental role here since their validity can be assumed for all metric theories of gravity where the local flat geometry of special relativity is  taken for granted.

\vspace{0.2cm}
\noindent{\it TEK law in Weyl integrable spacetime. -} The geometrical features of Weyl's theory consists of a 4-dimensional spacetime manifold $M$ with an extended symmetric connection  and a Lorentzian metric $g_{\mu\nu}$. In this geometry the connection is not a metric connection with respect to $g_{\mu\nu}$ and the covariant differentiation of the metric is given by
\begin{align} \label{nonmetricity}
 \stackrel{\mathcal{W}}{\nabla}_{\alpha}g_{\mu\nu} = A_{\alpha}g_{\mu\nu}\,, 
\end{align}
 where $\nabla$ with a superscript $\mathcal{W}$ indicate covariant derivatives in Weyl geometry. The quantities $A_{\alpha}$ are components in the coordinate basis $\{\partial_{\alpha}\}$ of some 1-form field $A$ defined on $M$.
 In what follow we will use the notation $\nabla$ and $\Gamma$ with a superscript letter $\mathcal{W}$ to indicate covariant derivatives and connections respectively in Weyl geometry, while the superscript ${\mathcal{R}}$ refers to Riemaniann geometry.

In the \emph{Weyl integrable spacetime}  $A$ is an exact 1-form, $A = d\phi$, where $\phi$ is a scalar function.  In the so-called Weyl frame, the gravitational interaction is now described by the pair [\,$g_{\mu\nu}(x^{\alpha}),\phi(x^{\alpha})\,]$.  This scalar field $\phi$ is of entirely geometric nature~\cite{salim}, and together with the metric $g_{\mu\nu}$ determine the WIST connection to be~\cite{Romero-Joel} 
\begin{equation}\label{Weyl-conec}
 \stackrel{\mathcal{W}}{\Gamma}\!^{\mu}_{\alpha\beta} = \stackrel{\mathcal{R}}{\Gamma}\!^{\mu}_{\alpha\beta} - \frac{1}{2}\left( \delta^{\mu}_{\alpha}\delta^{\nu}_{\beta} + \delta^{\mu}_{\beta}\delta^{\nu}_{\alpha} - g_{\alpha\beta} g^{\mu\nu} \right)\partial_{\nu}\phi\,,
\end{equation}
where $\stackrel{\mathcal{R}}{\Gamma}\!^{\mu}_{\alpha\beta} = g^{\mu\sigma}(\partial_{\alpha}g_{\sigma\beta} + \partial_{\beta}g_{\alpha\sigma} - \partial_{\sigma}g_{\alpha\beta})/2$ are the components of the Levi-Civita connection of $g_{\alpha\beta}$. 

The Weyl covariant divergence for a second order mixed tensor $T^{\alpha}\!_{\beta}$ is given by
\begin{equation}\label{Weyl-divergence}
   \stackrel{\mathcal{W}}{\nabla}_{\alpha}T^{\alpha}\!_{\beta} = \partial_{\alpha}T^{\alpha}\!_{\beta} +  \stackrel{\mathcal{W}}{\Gamma}\!^{\alpha}_{\mu\alpha}T^{\mu}\!_{\beta} -  \stackrel{\mathcal{W}}{\Gamma}\!^{\mu}_{\beta\alpha}T^{\alpha}\!_{\mu}\,.
\end{equation}
Substituting (\ref{Weyl-conec}) in (\ref{Weyl-divergence}), a straightforward calculation shows that
\begin{equation}\label{Weyl-div-Tab}
  \stackrel{\mathcal{W}}{\nabla}_{\alpha}T^{\alpha}\!_{\beta} =  \stackrel{\mathcal{R}}{\nabla}_{\alpha}T^{\alpha}\!_{\beta} - 2\,T^{\alpha}\!_{\beta}\,\partial_{\alpha}\phi + \frac{T^{\alpha}_{\alpha}}{2}\,\partial_{\beta}\phi\,,
\end{equation}
where $T^{\alpha}_{\alpha}$ is the trace of the tensor. Next let us consider that $T^{\alpha}_{\beta}$ is the matter source in Einstein's gravity. One may think, from (\ref{Weyl-div-Tab}), that in WIST geometry the conservation of the EMT is violated unless $\phi = constant$. However, this is not the case since the scalar field $\phi$ is an essential part of Weyl's geometry, and, as such, it should appear in any equation describing the behavior of the matter  in spacetime \cite{Romero-Joel}. 
With these considerations in mind, in what follows we consider $^{\mathcal{W}}{\nabla}_{\alpha}\,T^{\alpha}_{\beta} = 0$ in (\ref{Weyl-div-Tab}) for a self-gravitating static perfect fluid
whose energy-momentum tensor is given by (\ref{E1}). We thus obtain
\begin{eqnarray} \label{conservation-eq-1}
  u_{\beta} \stackrel{\mathcal{R}}{\nabla}_{\alpha}\left[(\rho + p)u^{\alpha} \right] +
  (\rho + p)u^{\alpha} \stackrel{\mathcal{R}}{\nabla}_{\alpha}\!u_{\beta} &    \nonumber  \\
  \mbox{}- \partial_{\beta}\,p - 2\,T^{\alpha}\!_{\beta}\,\partial_{\alpha}\phi + \frac{T^{\alpha}_{\alpha}}{2}\,\partial_{\beta}\phi = 0. &
\end{eqnarray}
Contracting this last equation with $u^{\beta}$ and substituting the result back in (\ref{conservation-eq-1}) we obtain, after some simplifications,
\begin{equation} \label{conservation-eq-2}
  u^{\alpha}\partial_{\alpha}u_{\beta} -  \stackrel{\mathcal{R}}{\Gamma}\!^{\mu}_{\alpha\beta} u^{\alpha}u_{\mu} +
  P^{\alpha}_{\beta}\left( \frac{\partial_{\alpha}p}{\rho + p} - \frac{1}{2}\,\partial_{\alpha}\phi \right) = 0
\end{equation}
where $P^{\alpha}_{\beta} = u^{\alpha}u_{\beta} - \delta^{\alpha}_{\beta}$ is the projector on the 3-space.

Now, in order to deduce the TEK law in WIST theory, we recall that for the static self-gravitating fluid described by the metric (\ref{M}), an observer at rest in the fluid has normalized 4-velocity $u^{\alpha}=\delta^{\alpha}_{0}/\sqrt{g_{00}}$ and $u_{\alpha}=\sqrt{g_{00}}\,\delta^0_{\alpha}$ and also that under static conditions $\partial_t\rho=\partial_t p=\partial_t\phi=0$. By taking this into account the conservation equation (\ref{conservation-eq-2}) reduces to
\begin{equation}\label{conservation-eq-4}
  \partial_i\ln\left[\,(\rho + p)\sqrt{g_{00}}\,\right] - \frac{1}{2}\,\partial_i\phi = \frac{\partial_i\rho}{\rho + p}\,.
\end{equation}

In this case, from (\ref{mu}) we have that $\rho + p = n(T\sigma + \mu)$ and substituting this into the EMT conservation law (\ref{conservation-eq-4}) for the general static configuration, we obtain
\begin{eqnarray}\label{conservation-eq-6}
  \partial_i\ln\left(T\sqrt{g_{00}}\right) - \frac{1}{2}\,\partial_i\phi =
  \frac{1}{\rho + p}\left[\partial_i\rho - \frac{\rho + p}{n}\,\partial_in \right.  & \nonumber  \\
 \mbox{} - \left. nT\partial_i\sigma - \frac{\rho + p}{\sigma + \mu/T}\,\partial_i\left( \frac{\mu}{T}\right) \right].  &
\end{eqnarray}
Now by taking into account  relation (\ref{GL1}), the above equation reduces to
\begin{equation}\label{TEKWmu}
  \partial_i\ln\left(T\sqrt{g_{00}}\right) - \frac{1}{2}\,\partial_i\phi + \frac{\partial_i(\mu/T)}{\sigma + \mu/T} = 0\,,
\end{equation}
which can be rewritten as:
\begin{equation}\label{TEKW}
 \partial_i\ln\left(T\sqrt{e^{-\phi}g_{00}}\right) + \frac{\mu}{T\sigma}\partial_i\ln\left(\mu\sqrt{e^{-\phi}g_{00}}\right) = 0\,.
\end{equation}
This is the Tolman-Ehrenfest-Klein's law for WIST when arbitrary values of the chemical potential are considered. By assuming $\mu = 0$, or even $\mu\sqrt{e^{-\phi}g_{00}}$ = constant (Klein's Law in WIST), it follows that    
\begin{equation}\label{TE2}
T\sqrt{e^{-\phi}g_{00}} = constant\,,
\end{equation}
which is the TE law in WIST theory. Conversely, we also see from (\ref{TEKW}) that the Klein law in WIST is also obtained when the TE law in WIST is taken for granted. In general, as should be expected, by comparing (\ref{TEKW}) and (\ref{TEKGRT})  we see that for $\phi$ = constant the TEK expression for GRT is recovered. 

\vspace{0.2cm}
\noindent{\it TEK law in ${f(R)}$ gravity.-}  The basics of $f(R)$ gravity is that in the Einstein-Hilbert action the Ricci scalar $R$ is substituted by a non-linear function $f(R)$, such that for $f(R)=R$ we recover the Einstein's general relativity.
However, in dealing with such theories two different variational approaches have been considered in the literature: (i) the metric, and (ii) the Palatini variational formulation. 

In the metric variational formalism the connection is assumed {\it a priori} to be the Levi-Civita one, thus the only independent variable in the action is the metric $g^{\mu\nu}$. In this case covariant derivatives must be taken with this connection. Although the dynamic equations in $f(R)$ gravity are very different  from GRT, the TEK result is immediate and will be announced here in the form of a theorem:

T1: {\it In $f(R)$ metric theory the TEK law is exactly the same of Einstein's gravity.} 

In light of the previous results [see Eq. (\ref{TEKGRT}) and (\ref{TEKW})], the logic and validity of the above theorem can be inferred  based on two complementary reasons:  (i) the deduction of TEK is independent of the field equations, and (ii) the covariant derivatives are exactly the same for $f(R)$  metric theory  and  Einstein's gravity (GRT). Nevertheless, in  Palatini's formulation the covariant derivatives depend on the specific function $f(R)$, thus the TEK theorem is modified as we shall see bellow. 

One of the  the main features of Palatini variational formulation is that in the gravitational action the connection and the metric are considered as independent fields while the matter action is assumed not to couple with the independent connections. In this case the Palatini connection coefficients are given by
\begin{equation} \label{Palatini-conect}
 \stackrel{P}{\Gamma}\!^{\mu}_{\alpha\beta} = \stackrel{\mathcal{R}}{\Gamma}\!^{\mu}_{\alpha\beta} + \frac{1}{2}\left( \delta^{\mu}_{\alpha}\delta^{\nu}_{\beta} + \delta^{\mu}_{\beta}\delta^{\nu}_{\alpha} - g_{\alpha\beta} g^{\mu\nu} \right)\partial_{\nu}\ln F
\end{equation}
where, as before, $\stackrel{\mathcal{R}}{\Gamma}\!^{\mu}_{\alpha\beta}$ are the components of the Levi-Civita connection of $g_{\alpha\beta}$ and $F = df/dR$.
Note that if $f(R)=R$ the two connections are equal and GRT is recovered.

The Palatini covariant divergence for a second order mixed tensor $T^{\alpha}_{\beta}$ is given by  
\begin{equation}\label{Palatini-divergence}
  \stackrel{P}{\nabla}_{\alpha}T^{\alpha}\!_{\beta} = \partial_{\alpha}T^{\alpha}\!_{\beta} + \stackrel{P}{\Gamma}\!^{\alpha}_{\mu\alpha}T^{\mu}\!_{\beta} - \stackrel{P}{\Gamma}\!^{\mu}_{\beta\alpha}T^{\alpha}\!_{\mu}\,.
\end{equation}
where $\nabla$ and $\Gamma$ with a superscript P denote covariant derivatives and connections in the Palatini setting. By replacing (\ref{Palatini-conect}) in the above equation, a straightforward calculation shows that
\begin{equation}\label{Palatini-div-Tab}
  \stackrel{P}{\nabla}_{\alpha}T^{\alpha}\!_{\beta} = \stackrel{\mathcal{R}}{\nabla}_{\alpha}T^{\alpha}\!_{\beta} + 2T^{\alpha}\!_{\beta}\partial_{\alpha}\ln F - \frac{T}{2}\,\partial_{\beta}\ln F
\end{equation}
where $T = T^{\alpha}_{\alpha}$ is the trace of $T^{\alpha\beta}$. Hence, following the similar reasoning to what we presented in WIST case, in what follows we put ${\nabla}^P_{\alpha}T^{\alpha}\!_{\beta} = 0$
in (\ref{Palatini-div-Tab}) for a static perfect fluid with energy-momentum tensor $T^{\alpha}\!_{\beta}$ as given by (\ref{E1}), thereby obtaining after some straightforward calculation
\begin{eqnarray}\label{conservation-Pal-1}
  \frac{1}{\sqrt{-g}}\,\partial_{\alpha}[ \sqrt{-g}\,(\rho + p)u^{\alpha}u_{\beta}] -
  (\rho + p)\stackrel{\mathcal{R}}{\Gamma}\!^{\mu}_{\alpha\beta}u^{\alpha}u_{\mu} & - \partial_{\beta}p   \nonumber  \\
  \mbox{} + 2(\rho + p)( u^{\alpha}u_{\beta} - \frac{1}{4}\,\delta^{\alpha}_{\beta})\partial_{\alpha}\ln F = 0. &
\end{eqnarray}
Next we assume that the self-gravitating fluid generates a static gravitational field described by the metric (\ref{M}), therefore the conservation equation (\ref{conservation-Pal-1}) reduces to
\begin{equation}\label{conservation-Pal-2}
  (\rho + p)\left[\partial_{\beta}\ln\sqrt{Fg_{00}}
   - 2(u^{\alpha}\partial_{\alpha}\ln F)\,u_{\beta}\right] + \partial_{\beta}p = 0.
\end{equation}
The $\beta=0$ component from (\ref{conservation-Pal-2}) yields $(\rho + p)\partial_t\ln F=0$, and considering $\rho + p \neq 0$, we have that $\partial_t\ln F =0$. This result is not generally valid for $f(R)$ gravity in the Palatini formulation, it is in fact a consequence of the particular spacetime (\ref{M}) used here. Therefore we obtain, after dividing by $\rho + p$, for the spatial components of (\ref{conservation-Pal-2}),
\begin{equation}\label{conservation-Pal-3}
  \partial_i\ln\left[\,(\rho + p)\sqrt{Fg_{00}}\,\right] = \frac{\partial_i\rho}{\rho + p}\,.
\end{equation}
In order to obtain the Tolman-Ehrenfest temperature law in Palatini $f(R)$ gravity we next use the Euler relation (\ref{mu}). The calculations are similar to those done  for WIST, so we present it here only briefly. Substituting (\ref{mu}) in (\ref{conservation-Pal-3}) we obtain, after some calculations, 
\begin{eqnarray}\label{conservation-Pal-5}
  \partial_i\ln\left(T\sqrt{Fg_{00}}\right) = \frac{1}{\rho + p}\left[\partial_i\rho - \frac{\rho + p}{n}\,\partial_in \right.  & \nonumber  \\
 \mbox{} - \left. nT\partial_i\sigma - \frac{\rho + p}{\sigma + \mu/T}\,\partial_i\left( \frac{\mu}{T}\right) \right].  &         
\end{eqnarray}
Now, by taking into account relation (\ref{GL1}) the above equation reduces to 
\begin{equation}\label{mu-TE-Palatini-law}
\partial_i\ln\left(T\sqrt{Fg_{00}}\,\right) + \frac{\partial_i (\frac{\mu}{T})}{\sigma + \frac{\mu}{T}} = 0\,,
\end{equation}
which can be rewritten as:
\begin{equation}\label{TEKF}
 \partial_i\ln\left(T\sqrt{Fg_{00}}\right) + \frac{\mu}{T\sigma}\partial_i\ln\left(\mu\sqrt{Fg_{00}}\right) = 0\,.
\end{equation}
This is the general form of Tolman-Ehrenfest-Klein's law for $f(R)$ gravity when arbitrary values of the chemical potential are considered. Again, by assuming $\mu=0$ or even $\mu\sqrt{F(R)g_{00}}$ = constant [Klein's Law in $f(R)$] it follows that    
\begin{equation}\label{TE3}
T\sqrt{F(R)g_{00}} = constant\,,
\end{equation}
which is the TE law in the Palatini version of $f(R)$ gravity. From (\ref{TEKF}) we also see that the Klein law in $f(R)$ is also obtained when the TE law is taken for granted. In general, we also see from  (\ref{TEKF}) and (\ref{TEKGRT}) that for $f(R) = R + const.$, the TEK expression for Einstein gravity is recovered [Cf. (\ref{TEKGRT})]. 

It is also interesting to note the similarity between the two connections given by relations (\ref{Palatini-conect}) and (\ref{Weyl-conec}), which become mathematically identical whether we make $F(R) \equiv \exp[-\phi(x)]$. However, we must keep in mind that WIST and Palatini $f(R)$ gravity have different physical base as well as different Lagrangian formulations. Without going into detail, in WIST the scalar field $\phi(x)$ already appears in the formulation of the gravitational action~\cite{Romero-Joel}, which must be varied in relation to the metric and to the field $\phi(x)$, while in Palatini formulation the fields considered as independent are the metric and the connections. Indeed, as shown in \cite{Sotiriou-Liberati}, $f(R)$ actions of gravity in Palatini formulation lead generically to theories with intrinsic nonmetricity. In this kind of theory the nonmetricity is induced by the form of $f(R)$ as ${\nabla}^{P}_{\alpha}g_{\mu\nu} = - \partial_{\alpha}(\ln{F})g_{\mu\nu}$. The only exception to this is the standard GRT linear function $f(R)$ = R + constant. Physically, however, the nonmetricity induced by Palatini $f(R)$ gravity is different from that presented in WIST. In \cite{Ghilencea}, for example, the author presents a comparative study of inflation in Weyl and Palatini quadratic gravity and shows how the different nonmetricities impact the predictions of some inflationary scenarios.  

\noindent{\it Discussion and Conclusion.-} In the present communication, we advanced a rigorous proof of the Tolman-Ehrenfest-Klein (TEK) thermodynamic theorem for some Riemannian and non-Riemannian geometries [see Eqs. (\ref{TEKW}), (\ref{TEKF}) and theorem 1]. The derivation presented here is as independent as possible of the properties of a specific medium or the physical state of matter. It means that  one of the main signatures of gravitation in thermodynamics which is the presence of gradients violating  the standard thermal equilibrium is heavily dependent on the underlying  gravitational theory. 

The main aspects of our proof are: (i) The TEK equations  (\ref{TEKGRT}), (\ref{TEKW}) and (\ref{TEKF}) are independent of the field equations for Riemannian and non-Riemannian geometries, but are heavily dependent on the covariant derivatives, and (ii) such unified relations uniting T, $\mu$ and the metric coefficient are valid for arbitrary expressions of the chemical potential.  As happens in the general relativity, the fluid is not restricted to blackbody radiation ($\mu=0$) and the approach followed here can naturally be extended for fluid mixtures. 

On experimental grounds, the direct verification of the general TEK law or some of its particular cases has  considerable physical interest, and, as such, deserves a closer scrutiny.  The results are based on the validity of Einstein's  equivalence principle, the apparatus of equilibrium relativistic thermodynamics, as well as on the underlying geometric structure of the spacetime.  Riemannian and non-Riemannian geometries satisfy different relations where the zeroth equilibrium thermodynamic condition are violated by the presence of gradients in the equilibrium states ($\Delta T\neq 0$,\, $\Delta\,{\mu} \neq 0$).  Thus, since the equivalence principle does not determine the  spacetime geometry, such results together are suggesting that its intrinsic geometric nature would also be revealed by a suitable crucial test in the thermal sector. 

Naturally, possible experimental tests may be  simplified by considering particular cases  of  TEK laws  for Weyl and Palatini $f(R)$ gravity theories. For example,  the  geometric nature of the spacetime can be tested by using the modified Tolman-Ehrenfest law with $\mu=0$ (photons) or more generally with $\mu/T$= constant [see Eqs. (\ref{TE2}) and (\ref{TE3})]. As far as we know, more than nine decades after the theoretical seminal paper of Tolman in Einstein gravity (1930), such a prediction has not been subjected to any experimental test. Probable, this happened because this astonishing effect is extremely tiny. In the weak field approximation, $g_{00} = 1+2\phi/c^{2}$, the gradient at the Earth surface with the radial position is $d\,lnT/dr= -g/c^{2} \simeq - 10^{-18}\,cm^{-1}$. Hence, it has   the same order of magnitude presented by the gravitational redshift (the apparent weight of photons) at Earth's surface which was tested only in the begin of the sixties through the so-called Pound-Rebka experiment \cite{PR}.  

The calculations presented here reinforce the interest to test this striking thermal-gravitational result delimiting the general validity of the standard zeroth thermodynamic law, not only as a sort of visual phenomenon, but a real thermal effect closely related with the intrinsic geometric nature of the spacetime. However,  apart the theoretical treatment,  such a challenging  experiment is not so simple at least in principle, remaining the question of its feasibility with the present day technology.     

\vskip 0.5cm {\bf Acknowledgments:} The authors are grateful to Carlos A. Romero for helpful correspondence. J.A.S.L. is partially supported by
CNPq (310038/2019-7), CAPES (88881.068485/2014), and
FAPESP (LLAMA Project No. 11/51676-9).

\end{document}